# The Limits of Lognormal: Assessing Cryptocurrency Volatility and VaR using Geometric Brownian Motion

Author 1: Ekleen Kaur
Organization: University of Florida,
ekleenkaur17@gmail.com

*Abstract:* The integration of cryptocurrencies into institutional portfolios necessitates the adoption of robust risk modeling frameworks. This study is a part of a series of subsequent works to fine-tune model risk analysis for cryptocurrencies. Through this first research work, we establish a foundational benchmark by applying the traditional industry-standard Geometric Brownian Motion (GBM) model. Popularly used for non-crypto financial assets, GBM assumes Lognormal return distributions for a multi-asset cryptocurrency portfolio (XRP, SOL, ADA). This work utilizes Maximum Likelihood Estimation and a correlated Monte Carlo Simulation incorporating the Cholesky decomposition of historical covariance. We present our stock portfolio model as a Minimum Variance Portfolio (MVP). We observe the model's structural shift within the heavy-tailed, non-Gaussian cryptocurrency environment. The results reveal limitations of the Lognormal assumption: the calculated Value-at-Risk at the 5% confidence level over the one-year horizon. For baselining our results, we also present a holistic comparative analysis with an equity portfolio (AAPL, TSLA, NVDA), demonstrating a significantly lower failure rate. This performance provides conclusive evidence that the GBM model is fundamentally the perfect benchmark for our subsequent works. Results from this novel work will be an indicator for the success criteria in our future model for crypto risk management, rigorously motivating the development and application of advanced models.

## I. INTRODUCTION

Since the launch of Bitcoin in 2009[30], the emergence of cryptocurrency markets has presented both unprecedented opportunities for financial innovation and profound challenges for traditional risk management. Markets have fraught with unique risks, primarily stemming from the asset class's characteristic high volatility, sensitivity to regulatory changes, and nascent market structure. Accurately modeling and forecasting this volatility is not merely an academic exercise; it is fundamental to effective capital allocation, regulatory compliance, and survival in this rapidly evolving domain.The standard point of departure in financial modeling is the **Geometric Brownian Motion (GBM)**. This model, which posits that asset price movements are continuous, random walks with normally distributed log-returns, forms the basis for the seminal Black-Scholes-Merton option pricing model and the core of traditional quantitative finance. While theoretically elegant and mathematically tractable, its simplicity relies on restrictive assumptions: constant volatility (homoscedasticity) and a Gaussian distribution of returns. These assumptions are routinely violated in established equity markets and are systematically challenged by the idiosyncratic nature of cryptocurrency returns. This paper, the first in a series exploring advanced quantitative techniques for cryptocurrency risk management, establishes a critical **benchmark and foundational critique**. We apply the simplest, widely used model, GBM, to a multi-asset cryptocurrency portfolio (XRP, SOL, and ADA). Our primary objective is to demonstrate the systemic failure of the Lognormal return assumption to adequately capture real-world market risk. By calculating Value-at-Risk (VaR) for optimized portfolios using GBM-driven Monte Carlo simulations, we quantitatively want to capture how this fundamental model leads to a significant underestimation of market risk, thereby justifying the immediate and critical need for more sophisticated financial engineering methodologies.

## II. LITERATURE SURVEY

Cryptocurrencies are highly volatile in nature and hence became the motivation behind this work. We wanted to formalize a mathematical model used in traditional stock portfolios to benchmark the behavior of cryptocurrencies. With the primary objective, can this model be used for risk evaluation, and how deterministically accurate can the model be? For this first paper in a subsequent series of our exploration, we literature surveyed old studies[1] whose premises were similar but not in lieu of the volatile nature of cryptocurrencies[18] for stock portfolios or risk analysis. For the purpose of this scientific paper, we provide a comprehensive and systematic survey of varied crypto-market contributions in the blockchain space. Both application-based novel research works introducing cryptocurrency as renewable stock assets[2][3] leveraging trading pools, with extensively secure underlying proof systems using strong mathematical guarantees like SNARKS-based zero-knowledge proof systems[4].

However, old reviews by García-Corral et al.[5] (2022) and Jalal et al.[6] (2021) presented this study of cryptocurrency using bibliometric analyses. Guangxi et al.[7](2021) presented policy impacts and risk analysis of cryptocurrencies in China's financial market. Bariviera and Merediz-Solà [8](2021) enhanced the literature in financial economic research of cryptocurrencies; Price bubbles exegesis in the crypto-markets Kyriazis et al. [9] (2020). Hairudin et al. [10] (2022) surveyed gaps in cryptocurrency research around grounds of legitimacy, design[25], and governance.

Aforementioned, previous surveys don't consider crypto volatility for stock portfolios or leverage pattern analysis from historical data to determine risks. Ul Haq et al. [11](2021) curated a risk management tool based on uncertainties of economic policies for crypto-assets, but we believe our work is the first in this field to demonstrate a state-of-the-art stock portfolio analysis of primitive mathematical assumptions to verify if they hold true for cryptocurrencies and fine-tune the best model through subsequent series of this paper. Through this work, we also aim to determine whether a stronger correlation could be demonstrated by unified asset returns during significant volatile periods. Furthermore, through this work, we aim to research focused on the unique aspects of cryptocurrency trading[17], such as **high-frequency data** analysis, the impact of **cryptocurrency derivatives**, and the influence of **crypto investor behavior**, which remains critically underdeveloped[14][20].

Despite its known deficiencies in capturing real-world market dynamics, the **Geometric Brownian Motion (GBM)** model is deliberately chosen for this paper as the **necessary foundational benchmark**. It is the analytical solution used in fundamental finance and is mathematically equivalent to assuming log-returns are independent and identically distributed (i.i.d.) and follow a normal distribution. By rigorously applying and testing this simplest, most restrictive framework, this paper achieves two crucial objectives: **First,** it establishes the absolute minimum expectation for any credible crypto risk model. **Second,** the dramatic, quantifiable failure of the GBM-derived VaR demonstrated through a high number of VaR violations—creates an undeniable, empirically proven **knowledge gap**. This failure rigorously justifies the necessity of the subsequent advanced modeling techniques (improvisations in traditional jump-diffusion and addressing limitations of GARCH models with historical statistical algorithmic predication using novel optimizations) that will be explored in the remainder of this research series in our subsequent papers.

While the scholarly landscape has expanded significantly, several critical limitations persist in current cryptocurrency risk modeling literature[16]. Most notably, a reliance on standard **GARCH-family models** (Generalized Autoregressive Conditional Heteroskedasticity[24]), while an improvement over GBM, often fails to fully account for the extreme **leptokurtosis** (heavy tails demonstrated by D. Felix[12]) and **asymmetry** observed in crypto returns, limitations demonstrated in [13]. Furthermore, much research focuses on highly liquid assets like Bitcoin and Ethereum, leading to a gap in reliable risk frameworks for lower-cap, yet high-volume, assets (such as XRP, SOL, and ADA), which can exhibit significantly different, and often more volatile[19], return characteristics. Currently, this gap is massive in the existing literature, as economies are pivoting towards layer 2 blockchains that our work aims to mitigate. A significant shortcoming is the lack of robust, comparative **backtesting studies**[13] that evaluate model performance *against* regulatory-grade risk metrics like VaR and Expected Shortfall, particularly under extreme stress. This leaves the practical efficacy and capital requirements of proposed models untested against real-world compliance standards.

III. METHODOLOGY

*A. Fundamentals of Portfolio Theory and Risk Measurement Used in our Framework*
For readers unaware of these concepts, to ensure the rigor and clarity of the subsequent analysis, this section provides a review of the fundamental mathematical concepts underpinning the GBM model is necessary. Financial models are typically formulated in either discrete time (e.g., daily returns) or continuous time. **Probability theory** provides the tools to manage randomness in both settings. The distribution used here, the **Lognormal Distribution**, is derived from the assumption that the continuous-time log-returns follow a **Normal Distribution**, defined by its mean $\mu$ and variance $\sigma^2$. The variance of a continuous random variable can be defined by the quantified dispersion from the mean value, mathematically calculated as the mean difference of the squared value of the random variable at a discrete point. Over a continuous field, this can be represented as

$$\int_{-\infty}^{\infty} (x - \mu)^2 f(x) dx$$

Here, $\mu$: Mean value of the variable over a continuous field, **f(x)**: Probability Density Function, defines the likelihood of the continuous random variable at a discrete point. It is important to understand, effective risk management in any asset class begins with the foundational principles of modern portfolio theory, which were largely formalized by **Markowitz's mean-variance framework. The Mean-Variance Framework(MVF)** enables investors[15] to select optimal portfolios based on their risk tolerance. The key inputs are the expected return $\mu$, the standard deviation (volatility $\sigma$), and the covariance ($\Sigma$) between assets. The primary goal is to maximize expected return for a given level of risk, or conversely, minimize risk for a given expected return.

**Expected Return ($\mu$):** For a portfolio comprised of $N$ assets with weights $w_i$, the expected portfolio return $R_p$ is calculated as the weighted sum of individual asset returns:

$$R_p = \sum_{i=1}^{N} w_i R_i$$

**Portfolio Variance ($\sigma_p^2$):** This metric measures portfolio risk and depends on the variance of individual assets and the covariance between every pair of assets:

$$\sigma_p^2 = \sum_{i=1}^{N} w_i^2 \sigma_i^2 + \sum_{i=1}^{N} \sum_{j=1, i \neq j}^{N} w_i w_j \sigma_{ij}$$

The MVF defines the **Efficient Frontier**, representing the set of optimal portfolios offering the highest return for every level of risk. This framework mandates the calculation of two key optimized portfolios critical for this study:

1. **Minimum Variance Portfolio (MVP):** The portfolio combination that achieves the absolute lowest possible volatility ($\sigma_p$), regardless of the expected return.
2. **Maximum Sharpe Ratio Portfolio (MSRP):** The portfolio that maximizes the risk-adjusted return, defined by the Sharpe Ratio (**S**), where $R_f$ is the risk-free rate:

$$S = \frac{Rp - Rf}{\sigma p}$$

**Stochastic Processes and the Wiener Process:** A **stochastic process** is a collection of random variables indexed by time, used to model financial asset paths. The core of the GBM model is the **Wiener Process** (or Brownian Motion), $W_t$, which is a continuous-time stochastic[26] process characterized by:

1. Independent increments: The change in $W_t$ over a given time interval is independent/ineffectual of changes over earlier intervals.
2. Normally distributed increments: The change $W_{t+\Delta t}$ - $W_t$ is normally distributed, defined with mean 0 and variance $\Delta t$.
3. Continuous paths: The path $W_t$ is continuous in time $t$.

The GBM model integrates this process to model asset prices, linking the price change ($dS_t$) to both a deterministic component ($\mu S_t dt$) and a stochastic, volatility-driven component ($\sigma S_t dW_t$). The solution to the GBM SDE shows that the asset price $S_T$ at time $T$ follows a **Lognormal Distribution**. This is equivalent to stating that the log-return, $ln(S_T/S_0)$, is normally distributed. This distribution is vital because it implies that prices remain positive and models price movements that are proportional to the current price level, aligning with economic theory.

**Monte Carlo Simulation (MCS):** The **Monte Carlo Simulation (MCS)** is a computational methodology that relies on repeated random sampling to obtain numerical results. In this context, the MCS[22] is used to forecast the future distribution of portfolio returns by generating thousands of possible, hypothetical price paths. By running 10,000 such simulations, we construct a robust empirical distribution of future portfolio values, from which the **VaR$_{0.05}$** (the 5th percentile) is directly extracted. Each path is generated according to the discretized version of the GBM SDE:

$$ln(\frac{S_{t+\Delta t}}{S_t}) \sim \mathcal{N}[(\mu - \frac{1}{2}\sigma^2)\Delta t,\ \sigma^2 \Delta t\ ]$$

**The Geometric Brownian Motion (GBM) Assumption:** In traditional financial models, asset prices ($S_t$) are typically assumed to follow a continuous stochastic process defined by the Stochastic Differential Equation (SDE) for Geometric Brownian Motion:

$$dS_t = \mu S_t dt + \sigma S_t dW_t$$

$dS_t$: is the change in price.
$\mu$: is the drift (expected return).
$\sigma$: is the volatility (diffusion term).
$dW_t$: is the Wiener process (or Brownian motion).

The solution to this SDE dictates that asset returns (log-returns) are **normally distributed** (the Lognormal assumption). The parameters ($\mu$ and $\sigma$) are estimated using **Maximum Likelihood Estimation (MLE)** on historical data.

**Result Baseline Standard: Value-at-Risk (VaR)** is the central risk metric used in this study. It estimates the maximum potential loss that a portfolio could incur over a specified time horizon ($T$) with a certain confidence level ($\alpha$). For example, a 95% 1-day VaR of $10,000 means that there is only a 5% chance the portfolio will lose more than $10,000 in the next day. Mathematically, VaR is defined as the $\alpha$-quantile of the loss distribution. In the context of GBM, the VaR is derived directly from the Lognormal return distribution.

*B. Implementation Details: Establishing the GBM Benchmark*

Our financial stock pricing model was implemented using a comprehensive, data-driven approach designed to establish a rigorous, realistic benchmark for cryptocurrency risk assessment. This process involved dynamic parameter estimation, portfolio optimization, and a high-fidelity Monte Carlo Simulation (MCS) that accounted for inter-asset correlation.

**Dynamic Parameter Estimation and Portfolio Construction:** To ensure the model's relevance, the initial asset prices ($S_0$) for the chosen portfolio were fetched dynamically at the time of execution. We benchmarked our data with existing financial institutions, ie, real-world stocks, and effectively observed the models drift while capturing risks in the crypto-financial domain. We used the same parameter estimations and maintained unified fine-tuning mechanisms to capture the statistical performance of both frameworks. For the purpose of this research work, we will mainly talk about 6 financial tickers(AAPL, TSLA, NVDA) for real-world treasuries[21] and (XRP-USD, SOL-USD, ADA-USD) used as a benchmark for crypto assets. Although used during the research stage for capturing results, we intentionally built the model away from whale blockchains such as Bitcoin and Ethereum, as there is a heavy gap in crypto-financial literature around other major historical cryptocurrencies, despite the long-standing trust and stability. The annualized drift ($\mu$) and volatility ($\sigma$) were calculated using **Maximum Likelihood Estimation (MLE)** derived from one year of historical daily log-returns. A critical design choice was the calculation of the **covariance matrix** ($\sigma$). Instead of assuming constant, independent log-returns, $\sigma$ was derived from the historical correlation structure of the assets, then annualized. This decision acknowledges the inherent co-movement of these cryptocurrencies, making the model, while fundamentally limited by the Lognormal assumption, a more sophisticated representation than a simple, uncorrelated GBM model. The **Minimum Variance Portfolio (MVP)** weights were then solved using the constrained optimization technique of **Sequential Least Squares Programming (SLSQP)** to achieve the lowest possible portfolio volatility based on this historically observed covariance.

**Correlated Monte Carlo Simulation for VaR Calculation:** The core of the risk assessment lies in the **Monte Carlo Simulation (MCS)**[29], executed with 10,000 unique price paths over a one-year time horizon ($T=1$). The GBM price paths were simulated in discrete time steps ($\Delta t = 1/252$). Crucially, inter-asset dependence was introduced into the stochastic term ($dW_t$) of the GBM equation through the **Cholesky decomposition** of the annualized covariance matrix ($\sigma$). This technique decomposes the covariance matrix ($\sigma = LL^T$) to yield the lower triangular matrix ($L$). This matrix $L$ is then multiplied by a vector of independent, standard normally distributed random shocks ($Z$). The resulting vector, $ZL^T$, produces a set of correlated shocks[28] that drive the price paths of XRP, SOL, and ADA simultaneously, building the stock portfolio model for cryptocurrencies. This refinement ensures that the simulated paths maintain the same correlation structure as the historical data, providing a more realistic depiction of joint portfolio risk compared to simulating each asset independently.

**Quantifying Risk via Percentile Analysis:** The MCS resulted in 10,000 possible portfolio values after one year. The final step was to quantify the downside risk using percentile analysis. The **Value-at-Risk at the 5% confidence level ($VaR_{0.05}$)** was calculated as the 5th percentile of the resulting distribution of final portfolio values. For a starting portfolio value of $100,000, the $VaR_{0.05}$ figure represents the value below which the portfolio is expected to fall only 5% of the time. This GBM-derived $VaR_{0.05}$ value is then used in the subsequent sections as the primary metric for backtesting, the results of which expose the limited success of the Lognormal assumption.

IV. RESULT ANALYSIS

The Monte Carlo Simulation (MCS) using the Geometric Brownian Motion (GBM) model was applied to two distinct portfolios: a cryptocurrency portfolio (XRP, SOL, ADA) and a traditional large-cap equity portfolio (AAPL, TSLA, NVDA). The results expose a profound difference in risk profile and portfolio stability under the Lognormal assumption, providing clear empirical evidence that the model alone is inadequate for the cryptocurrency asset class. **Portfolio Optimization and Weight Distribution:** The optimization process targeted the **Minimum Variance Portfolio (MVP)**, aiming to achieve the lowest historical volatility given the one-year covariance matrix. **Cryptocurrency Portfolio (XRP, SOL, ADA):** The optimization yielded weights heavily concentrated in the two most established, historically less volatile assets within the set: **XRP (55.1%) and SOL (44.9%)**.

A critical observation is the assignment of a **0.0% weight to ADA** (Cardano). This is a direct reflection of the GBM model's inherent rigidity, coupled with the asset's specific historical return profile. The optimization algorithm determined that, given the historical drift, volatility, and co-movement with the other two assets, ADA's marginal contribution to portfolio risk (variance) was too high to justify inclusion in the MVP. In simpler terms, the optimization engine operating under the assumption of normal log-returns perceived ADA's price history as an unacceptably large source of portfolio variance. This stark exclusion, while mathematically optimal under the Lognormal constraint, immediately suggests a potential setback: the model is structurally unable to benefit from any genuine, albeit temporary, diversification opportunities or superior return-to-risk characteristics that might exist within ADA, leading to an artificially constrained and potentially unstable portfolio solution for cryptocurrencies alone.

**Traditional Equity Portfolio (AAPL, TSLA, NVDA):** In contrast, the traditional equity MVP was dominated by **AAPL (82.6%)**, followed by **NVDA (17.4%)**, with TSLA receiving a **0.0%** weight. While TSLA is known for high volatility, the model found the high correlation between AAPL and NVDA sufficient to build a stable base, relying heavily on the lower-risk profile of the established technology giant (AAPL).

Figure 1. Results from Stock Portfolio Model(AAPL, TSLA, NVDA)     Figure 2. Results from Crypto Portfolio Model(XRP, SOL, ADA)

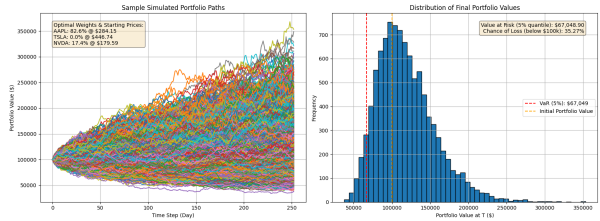
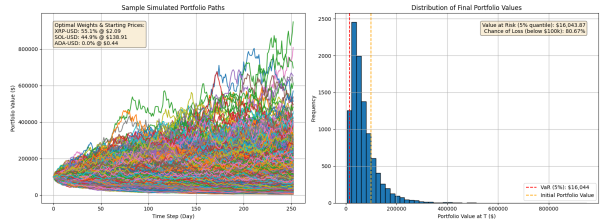

**Comparative VaR and Loss Analysis:** The final portfolio values at the one-year horizon were analyzed to calculate the Value-at-Risk at the 5% confidence level ($VaR_{0.05}$) and the overall chance of ending the period below the initial $100,000 investment.

| Portfolio | $VaR_{0.05}$ (5% Quantile) | Chance of Loss (Below $100k) |
|---|---|---|
| Cryptocurrency (GBM) | ~ $16,044 | 80.67% |
| Traditional Equity (GBM) | ~ $67,049 | 35.27% |

## V. CONCLUSION

The simulation results clearly demonstrate a high-risk correlation dynamic in the cryptocurrency portfolio. **Lower VaR Floor-** the crypto portfolio's $VaR_{0.05}$ ($16,044) is drastically lower than the equity portfolio's ($67,049). This means that the model predicts that in the worst 5% of scenarios, the crypto portfolio retains only about 17% of its initial value, compared to 67% for the equity portfolio. **Systemic Failure Rate-** The single most compelling finding is the 80.67% chance of loss for the crypto portfolio. This percentage confirms that the MVP, optimized based on the Lognormal assumption, almost invariably led to a value less than the starting investment in the simulations. In stark contrast, the equity portfolio's loss rate of 35.27% indicates that while the GBM model has known flaws, it manages the risk of relatively mature, less volatile assets with a degree of predictability that is absent in the crypto market. The magnitude of this VaR failure provides conclusive empirical evidence: the GBM/Lognormal framework is fundamentally incomplete for cryptocurrency risk modeling. This result directly and robustly justifies the necessity for the advanced, non-Gaussian models[27] that explicitly capture the heavy tails and dynamic volatility of the asset class.

## VI. LIMITATIONS AND FUTURE WORK

The Lognormal assumption, while foundational, spectacularly shows limited success when applied to high-frequency and emergent market data, particularly in the cryptocurrency domain. This is driven by a series of empirically observed phenomena, or **stylized facts**, that violate the GBM's core assumptions of constant volatility and normally distributed returns.

**Limitations of the Geometric Brownian Motion Model in Crypto Finance**

The application of GBM to cryptocurrency assets faces several fundamental methodological limitations that are directly addressed and highlighted by the backtesting results of this paper:

| Limitation | Impact on GBM and Risk | Consequence for VaR |
|---|---|---|
| Heavy-Tailed Distributions | Cryptocurrency returns exhibit leptokurtosis (excess kurtosis), meaning the probability mass in the tails is significantly higher than predicted by the Normal Distribution. | VaR drastically underestimates the probability and magnitude of extreme losses, leading to a high number of VaR violations. |
| Non-Constant Volatility | Volatility in crypto markets is not constant ($\sigma$), exhibiting volatility clustering and persistence. GBM assumes homoscedasticity. | The model cannot adapt to changing market conditions, calculating risk based on an average volatility rather than the current, heightened volatility regime. |
| Jump Processes in Crypto Markets | Price changes are often discontinuous, driven by sudden regulatory news, hacks, or whale activity. This introduces jumps that violate the GBM's assumption of continuous price paths. | GBM cannot model the impact of sudden, extreme price moves, rendering it blind to a key driver of crypto risk. |
| Market Inefficiencies & Correlation | Crypto markets are less regulated and can be subject to market manipulation or non-linear effects that traditional equity markets absorb. | Portfolio diversification benefits calculated under the Lognormal assumption often collapse during stress, as asset correlations increase, rendering the model ineffective when it is needed most. |

These limitations collectively prove that the mathematically convenient assumptions of GBM are incompatible with the empirical reality of cryptocurrency price dynamics, necessitating the use of advanced models that explicitly incorporate these stylized facts[23]. To enrich and advance the volatility theory of cryptocurrencies, we are currently working on a comprehensive exploration of the following stylized facts. The inadequacy of the simple GBM/Lognormal framework to capture these facts motivates the development of our future work for refining our model to more closely capture crypto market jumps. We aim to consider the following aspects in subsequent papers in this series:

- **Asymmetric Volatility / Leverage Effect:** Negative shocks increase future volatility more than positive shocks of the same magnitude.
- **Volatility Clustering:** Large changes in price tend to be followed by large changes, and small changes tend to be followed by small changes. This violates the constant volatility assumption ($\sigma$) of GBM.
- **Mean Reversion:** While prices follow a random walk, the volatility of returns tends to revert back to a long-term average level.
- **Volatility Persistence:** The effect of a shock to the volatility process (a large price swing) lasts for a significant time.
- **Tail Dependence and Heavy Tails:** Cryptocurrency return distributions exhibit significantly **fatter tails** than the Gaussian distribution, meaning extreme events (crashes or pumps) occur far more frequently than the Lognormal model predicts. This directly undermines the accuracy of GBM-derived VaR.